\documentclass[conference]{IEEEtran}
\usepackage[noadjust]{cite}
\usepackage{amsmath}
\usepackage{amsfonts}
\usepackage{graphicx}
\usepackage{float}
\usepackage{subfigure}
\usepackage{epsfig}
\usepackage{amsmath}
\usepackage{theorem}
\usepackage{amsmath}
\usepackage{mathrsfs}
\usepackage{amsfonts}
\usepackage{amssymb}
\usepackage{mathrsfs}
\usepackage{euscript}
\usepackage{multirow}
\usepackage{cite}
\usepackage{url}
\usepackage{color}
\usepackage{algorithm}
\usepackage{algorithmic}
\usepackage{flushend}
\def\BibTeX{{\rm B\kern-.05em{\sc i\kern-.025em b}\kern-.08em
    T\kern-.1667em\lower.7ex\hbox{E}\kern-.125emX}}

\begin{document}
\title{Rate Balancing in Full-Duplex MIMO Two-Way Relay Networks}
\author{Erfan Khordad $^{\dag}$,~Ata Khalili $^{\ddag}$,~and Soroush Akhlaghi $^{\ddag}$\\$^{\dag}$~Department of Engineering, Macquarie University,~Australia \\$^{\ddag}$~Department of Engineering, Shahed University,~Iran }
\maketitle
\begin{abstract}
Maximizing the minimum rate for a full-duplex multiple-input multiple-output~(MIMO) wireless network encompassing two sources and a two-way~(TW) relay operating in a two-hop manner is investigated. To improve the overall performance, using a zero-forcing approach at the relay to suppress the residual self-interference arising from full-duplex~(FD) operation, the underlying max-min problem is cast as an optimization problem which is non-convex. To circumvent this issue, semi-definite relaxation technique is employed, leading to upper and lower bound solutions for the optimization problem. Numerical results verify that the upper and lower bound solutions closely follow each other, showing that the proposed approach results in a close-to-optimal solution. In addition, the impact of residual self-interference upon the overall performance of the network in terms of the minimum rate is illustrated by numerical results, and for low residual self-interference scenarios the superiority of the proposed method compared to an analogous half-duplex~(HD) counterpart is shown.
\end{abstract}

\IEEEpeerreviewmaketitle
\section{Introduction}
Spectral efficiency is regarded as one of the most indispensable factors in wireless communications, especially in recent years which spectrum resources are at a premium. FD operation can be a promising solution to satisfy this demand. More precisely, FD communication systems have the potential to double the spectral efficiency since they enable simultaneous transmitting and receiving data over the same frequency band.

Relaying is another method to establish high-quality wireless networks~\cite{Aazhang}. Employing relay-aided communications increases network coverage, diversity gain, and also spectral efficiency. One-way~(OW) and TW relaying are two strategies chiefly deployed in cooperative networks. OW relaying requires two communications phases. In the first phase, the transmitter sends its signal to the relay, then the relay based on its certain scheme processes the received signal and sends it to the receiver. TW relaying also consists of two communications phases; however, unlike OW relaying, two transceivers exchange their data owing to the bi-directional connection established between them by the relay~\cite{Shahbazpanahi}.

FD operation can be employed at the relay to make the relaying strategy more spectral efficient and enable one-phase date exchange between transceivers. In the literature, FD relaying has been investigated for both of the OW and TW strategies. The major concern in FD relay networks is the self-interference~(SI) of the relay, degrading its performance extensively. However, the possibility of effective communications in the presence of SI is justified in~\cite{Riihonen}. Consequently, \cite{RiihonenWerner2} developed a method to change opportunistically between FD and HD OW relaying modes. Assuming FD OW relays, in \cite{Krikidis}, an optimal relay selection method using global channel state information~(CSI) knowledge and several relay selection schemes by using partial CSI are studied. On the other hand, FD TW relaying increases the data rate twice that of the HD relaying, thereby enhancing the spectral efficiency. In \cite{Hcui}, considering multiple FD TW relays, the optimal relay selection method to maximize the signal-to-interference plus noise ratio~(SINR) is investigated.

Using spatial domain signal processing can also be an effective solution to alleviate the detrimental effect of SI. In \cite{RiihonenWichman3}, some techniques to mitigate SI in FD MIMO OW relay networks are investigated. In~\cite{Js-rdffdmimoafrs}, assuming FD MIMO OW relaying, the rate maximization problem for the case when the relay's amplification matrix is of rank one is converted to an unconstrained problem solved by the gradient method. Also, an approach to deal with the issue resulting from SI is proposed in~\cite{Js-rdffdmimoafrs} for the general case when the relay's amplification matrix is not of rank one.

In~\cite{GZheng}, the beamforming matrix used at the relay and the power allocation of the sources are jointly optimized for a FD TW MIMO relay network. In particular, the achievable rate region and the sum rate maximization are studied in~\cite{GZheng}. In~\cite{bdfFDTWafMR} considering a FD MIMO TW relay network with imperfect cancellation of SI, and by minimizing the mean square error under a relay transmit power constraint, the relay beamforming matrix and receive beamforming matrices at sources are designed. A beamforming design that reserves a fraction of SI is proposed in~\cite{jRUbdiafutwrch} for a FD TW relay channel with multiple antennas, and the local optimum for relay beamformers and the global optimum for user beamformers are obtained. \cite{tDfMIMOfdtWRn} considered loopback SI at all nodes of a FD MIMO TW relay network, and designed precoding for all nodes to reduce the residual loopback SI, along with deriving a recursive expression showing the cumulative impact of the residual loopback SI.

To the best of the authors' knowledge, the problem of establishing fairness in terms of the rate for the two sources in a MIMO FD TW relay network has not yet been investigated. Therefore, in this paper, considering a MIMO FD TW relay network and assuming a zero-forcing constraint at the relay to eliminate the residual self-interference~(RSI), the problem of maximizing the minimum rate of transmitting ends is investigated to enhance the fairness. To this end, the underlying problem is formulated as an optimization problem which is not convex in general. Hence, incorporating the so-called semi-definite relaxation~(SDR) technique, the original problem is converted into a semi-definite programming~(SDP) problem, where upper and lower bounds associated with the original problem are numerically derived. Numerical results illustrate that upper and lower bound solutions closely follow each other, showing that a close-to-optimal solution is derived. We also use the approach adopted in~\cite{maxminSINRref} as a baseline scheme for the performance evaluation of our proposed method. It is worth mentioning that, considering a HD MIMO TW relaying system and employing the Dinkelbach algorithm, the relay's beamforming matrix maximizing the minimum received SNR of two sources is obtained in~\cite{maxminSINRref}.

The rest of the paper is organized as follows. The system model is presented in Section~\ref{system}. In Section~\ref{problem}, the optimization problem is formulated. Numerical results are presented in Section \ref{simulation}. Section~\ref{complexity} compares the computational complexity of the proposed approach with that of presented in~\cite{maxminSINRref}. Finally, the conclusion is drawn in Section~\ref{conclusion}.

\section{System Model}\label{system}
We consider a wireless network including two FD sources $S_i$ where $i\in \{1,\,2\}$, and an amplify-and-forward FD TW relay $R$ as it is shown in Fig.~1.~Owing to the path loss, shadowing and intense multipath impacting negatively on the wireless channels, it is assumed that there is no direct link between sources and they are obliged to use the relay in order to establish the connection. In this network, the CSI is globally available among all nodes. Moreover, $S_i$ is equipped with two antennas; one for transmitting and another for receiving. In addition, $R$ has $M\geq4$ even antennas; $N$ transmit antennas and $N$ receive antennas where $N=\frac{M}{2}$. We also assume that most of SI is canceled out, and merely RSI channels are assumed between the receive and the transmit antennas of each node~\cite{RiihonenWichman3}. In the aforementioned structure, $S_1$ and $S_2$ aim at exchanging information through using $R$, where their transmit signals to $R$ are denoted by $x_i$ with power $p_i$. Since all nodes are operating in the FD mode, they simultaneously receive and transmit signals. Therefore, at the time slot $k$, $R$'s received signal can be written as follows
\begin{equation}\label{receiveR}
{{\mathbf{y}}_{R}}\left( k \right)={{\mathbf{f}}_{1R}}{{x}_{1}}\left( k \right)+{{\mathbf{f}}_{2R}}{{x}_{2}}\left( k \right)+{{\mathbf{H}}_{RR}}{{\mathbf{x}}_{R}}\left( k \right)+{{\mathbf{n}}_{R}}\left( k \right),
\end{equation}
where ${{\mathbf{f}}_{iR}}\in {{\mathbb{C}}^{N\times 1}}$ is the quasi-static channel gain vector of the $S_i$-$R$ link, ${\mathbf{H}}_{RR}$ is the RSI channel gain matrix between $R$'s transmit and receive antennas as it is assumed in~\cite{GZheng}, and ${{\mathbf{n}}_{R}}\sim\mathcal{CN}\left( \mathbf{0},\sigma _{R}^{2}\mathbf{I} \right)$ is the received noise at $R$.

Moreover, the $R$'s transmit signal at the time slot $k$ is given by
\begin{equation}\label{R_transmit_signal}
{{\mathbf{x}}_{R}}\left( k \right)=\mathbf{W}{{\mathbf{y}}_{R}}\left( k-1 \right),
\end{equation}
where $\mathbf{W}$ is the $R$'s beamforming matrix which meets the zero-forcing constraint~(ZFC), i.e., $\mathbf{W}{{\mathbf{H}}_{RR}}={{\mathbf{0}}_{N\times N}}$, eliminating the RSI at the node $R$. Consequently, $R$'s power can be computed as
\begin{equation}\label{R_power}
\begin{aligned}
{{{p}}_{{R}}}&={E}\left\{ {\mathbf{x}_R^H}\mathbf{x}_R \right\}= {Tr}\left( \mathbf{W}{E}\left\{{{\mathbf{y}_R\mathbf{y}_R^H}}\right\}{{\mathbf{W}}^{{H}}} \right)\\
&=\,{Tr}\left( \mathbf{W}{{\mathbf{L}}_R}{{\mathbf{W}}^{{H}}} \right)
\overset{\left( \text{a} \right)}{\mathop{=}}\,{vec}{{\left( \mathbf{L}_R^{{T}}\mathbf{I}{{\mathbf{W}}^{{T}}} \right)}^{{T}}}{vec}\left( {{\mathbf{W}}^{{H}}} \right)\\
&\overset{\left( \text{b} \right)}{\mathop{=}}\,{vec}{{\left( \mathbf{I} \right)}^{{T}}}{{\left( \mathbf{W}\otimes \mathbf{L}_R^{{T}} \right)}^{{T}}}{vec}\left( {{\mathbf{W}}^{{H}}} \right)\\
&
\overset{\left( \text{c} \right)}{\mathop{=}}\,{vec}{{\left( \mathbf{I} \right)}^{{T}}}\left( {{\mathbf{W}}^{{T}}}\otimes \mathbf{L}_R \right){vec}\left( {{\mathbf{W}}^{{H}}} \right)\\& \overset{\left( \text{d} \right)}{\mathop{=}}\,{vec}{{\left( {{\mathbf{W}}^{{T}}} \right)}^{{T}}}\left( \mathbf{I}\otimes \mathbf{L}_R \right){vec}\left( {{\mathbf{W}}^{{H}}} \right)\\
&\overset{\left( \text{e} \right)}{\mathop{=}}\,{{\mathbf{w}}^{{H}}}\left( \mathbf{I}\otimes \mathbf{L}_R \right)\mathbf{w}
={{\mathbf{w}}^{{H}}}\mathbf{C}_1\mathbf{w},
\end{aligned}
\end{equation}
where in (\ref{R_power}), (a), (b), (c) and (d) come from ${Tr}\left( {{\mathbf{A}}^{{T}}}\mathbf{B} \right)={vec}{{\left( \mathbf{A} \right)}^{{T}}}{vec}\left( \mathbf{B} \right)$, ${vec}(\textbf{ABC}) = (\textbf{C}^{{T}} \otimes \textbf{A}){vec}(\textbf{B})$, ${{\left( \mathbf{A}\otimes \mathbf{B} \right)}^{{T}}}=\left( {{\mathbf{A}}^{{T}}}\otimes {{\mathbf{B}}^{{T}}} \right)$ and ${vec}{{\left( \mathbf{I} \right)}^{{T}}}\left( {{\mathbf{A}}^{{T}}}\otimes \mathbf{B} \right)={vec}{{\left( {{\mathbf{A}}^{{T}}} \right)}^{{T}}}\left( \mathbf{I}\otimes \mathbf{B} \right)$, respectively. Also, we have $\mathbf{L}_R={E}\left\{{{\mathbf{y}_R\mathbf{y}_R^H}}\right\}$ and $\mathbf{C}_1=\mathbf{I}\otimes \mathbf{L}_R$, and defining the beamforming vector $\mathbf{w}$ as $\mathbf{w}=\text{vec}\left( {{\mathbf{W}}^{\text{H}}} \right)$ leads to (e).

By rewriting the ZFC as $\mathbf{H}_{RR}^{H}{{\mathbf{W}}^{H}}\mathbf{I}={{\mathbf{0}}_{N\times N}}$ and using (b), one can convert the ZFC to $\left( \mathbf{I}\otimes \mathbf{H}_{RR}^{H} \right)vec\left( {{\mathbf{W}}^{H}} \right)={{0}_{{{N}^{2}}\times 1}}$ which gives ${{\mathbf{w}}^{H}}{{\mathbf{C}}_{2}}\mathbf{w}=0$, where ${{\mathbf{C}}_{2}}=\mathbf{I}\otimes \mathbf{H}_{RR}^{H}$.

On the other hand, $S_i$ receives the following signal at the $k^{th}$ time slot
\begin{equation}\label{source_received_signal}
{{y}_{i}}\left( k \right)=\mathbf{f}_{Ri}^{H}{{\mathbf{x}}_{R}}\left( k \right)+{{{f}}_{ii}}{{x}_{i}(k)}+{{n}_{i}}\left( k \right),
\end{equation}
where ${{\mathbf{f}}_{Ri}}\in {{\mathbb{C}}^{N\times 1}}$ is a quasi-static channel gain vector of the $R$-$S_i$ link, $f_{ii}$ is the RSI channel gain vector between $S_i$'s transmit and receive antennas and ${{{n}}_{i}}\sim\mathcal{CN}\left( {0},\sigma _{i}^{2} \right)$ is the received noise at $S_i$.

Using (\ref{receiveR}), (\ref{R_transmit_signal}) and ZFC, (\ref{source_received_signal}) can be rewritten as follows
\begin{equation}\label{sourcereceivedsignal2}
\begin{aligned}
{{y}_{i}}\left( k \right)&=\mathbf{f}_{Ri}^{H}\mathbf{W}{{\mathbf{f}}_{iR}}{{x}_{i}(k-1)}+\mathbf{f}_{Ri}^{H}\mathbf{W}{{\mathbf{f}}_{\left( 3-i \right)R}}{{x}_{\left( 3-i \right)}(k-1)}\\&+\mathbf{f}_{Ri}^{H}\mathbf{W}{{\mathbf{n}}_{R}(k-1)}+{{\mathbf{f}}_{ii}}{{x}_{i}(k)}+{{n}_{i}}\left( k \right).
\end{aligned}
\end{equation}
Note that $R$ sends $\mathbf{W}$ to $S_i$ at the start of each transmission block; in addition, $S_i$ has access to CSI of ${{\mathbf{f}}_{iR}}$ and ${{\mathbf{f}}_{Ri}}$; thus, the term $\mathbf{f}_{Ri}^{H}\mathbf{W}{{\mathbf{f}}_{iR}}{{x}_{i}(k-1)}$ in (\ref{source_received_signal}) is known for $S_i$ and can be simply canceled out. As a result, $S_i$'s received SINR can be computed as follows
\begin{equation}\label{SINR_i}
SIN{{R}_{i}}=\frac{{{p}_{\left( 3-i \right)}}{{\left| \mathbf{f}_{Ri}^{H}\mathbf{W}{{\mathbf{f}}_{\left( 3-i \right)R}} \right|}^{2}}}{{{\left\| \mathbf{f}_{Ri}^{H}\mathbf{W} \right\|}^{2}}+{{p}_{i}}{{\left| {{\mathbf{f}}_{ii}} \right|}^{2}}+\sigma _{i}^{2}}.
\end{equation}
In order to rewrite (\ref{SINR_i}) in terms of the vector $\mathbf{w}$, the term ${{\left| \mathbf{f}_{Ri}^{H}\mathbf{W}{{\mathbf{f}}_{\left( 3-i \right)R}} \right|}^{2}}$ in the numerator of (\ref{SINR_i}) can be computed as
\begin{equation}\label{termsofSINR}
\begin{aligned}
&{{\left| \mathbf{f}_{Ri}^{H}\mathbf{W}{{\mathbf{f}}_{\left( 3-i \right)R}} \right|}^{2}}=Tr\left( \mathbf{f}_{Ri}^{H}\mathbf{W}{{\mathbf{f}}_{\left( 3-i \right)R}}\mathbf{f}_{\left( 3-i \right)R}^{H}{{\mathbf{W}}^{H}}{{\mathbf{f}}_{Ri}} \right)\\
&\overset{(\text{a})}{\mathop{=}}Tr\left( \mathbf{W}{{\mathbf{f}}_{\left( 3-i \right)R}}\mathbf{f}_{\left( 3-i \right)R}^{H}{{\mathbf{W}}^{H}}{{\mathbf{f}}_{Ri}}\mathbf{f}_{Ri}^{H} \right)\\&\overset{(\text{b})}{\mathop{=}}ve{{c}^{H}}\left( {{\mathbf{W}}^{H}} \right)vec\left( {{\mathbf{f}}_{\left( 3-i \right)R}}\mathbf{f}_{\left( 3-i \right)R}^{H}{{\mathbf{W}}^{H}}{{\mathbf{f}}_{Ri}}\mathbf{f}_{Ri}^{H} \right)\\
&\overset{(\text{c})}{\mathop{=}}{{\mathbf{w}}^{H}}\left( \mathbf{f}_{Ri}^{*}\mathbf{f}_{Ri}^{T}\otimes {{\mathbf{f}}_{\left( 3-i \right)R}}\mathbf{f}_{\left( 3-i \right)R}^{H} \right)vec\left( {{\mathbf{W}}^{H}} \right)\\
&={{\mathbf{w}}^{H}}{{\mathbf{H}}_{i}}\mathbf{w},
\end{aligned}
\end{equation}
where (a), (b) and (c) come from $Tr\left( {{\mathbf{a}}^{H}}\mathbf{b} \right)=Tr\left( \mathbf{b}{{\mathbf{a}}^{H}} \right)$, $Tr\left( {{\mathbf{A}}^{H}}\mathbf{B} \right)=ve{{c}^{H}}\left( \mathbf{A} \right)vec\left( \mathbf{B} \right)$ and $vec\left( \mathbf{ABC} \right)=\left( {{\mathbf{C}}^{T}}\otimes \mathbf{A} \right)vec\left( \mathbf{B} \right)$ respectively, and also ${\mathbf{H}}_{i}=\mathbf{f}_{Ri}^{*}\mathbf{f}_{Ri}^{T}\otimes {{\mathbf{f}}_{\left( 3-i \right)R}}\mathbf{f}_{\left( 3-i \right)R}^{H}$. By the same token, we have ${{\left\| \mathbf{f}_{Ri}^{H}\mathbf{W} \right\|}^{2}}={{\mathbf{w}}^{H}}{{\mathbf{F}}_{i}}\mathbf{w}$ where ${{\mathbf{F}}_{i}}=\mathbf{f}_{Ri}^{*}\mathbf{f}_{Ri}^{T}\otimes \mathbf{I}$. Therefore, the received rate at the node $S_i$ can be written as follows
\begin{equation}\label{rewrite_SINRi}
{{\Re }_{i}}={{\log }_{2}}\left( 1+\frac{{{\mathbf{w}}^{H}}{{\mathbf{H}}_{i}}\mathbf{w}}{{{\mathbf{w}}^{H}}{{\mathbf{F}}_{i}}\mathbf{w}+{{\phi }_{i}}} \right),
\end{equation}
where ${{\phi }_{i}}={{p}_{i}}{{\left| {{\mathbf{f}}_{ii}} \right|}^{2}}+\sigma _{i}^{2}$.
\begin{figure}[t]\label{system_model}
\centering
 \includegraphics[width=8.00cm,height=3.2500cm]{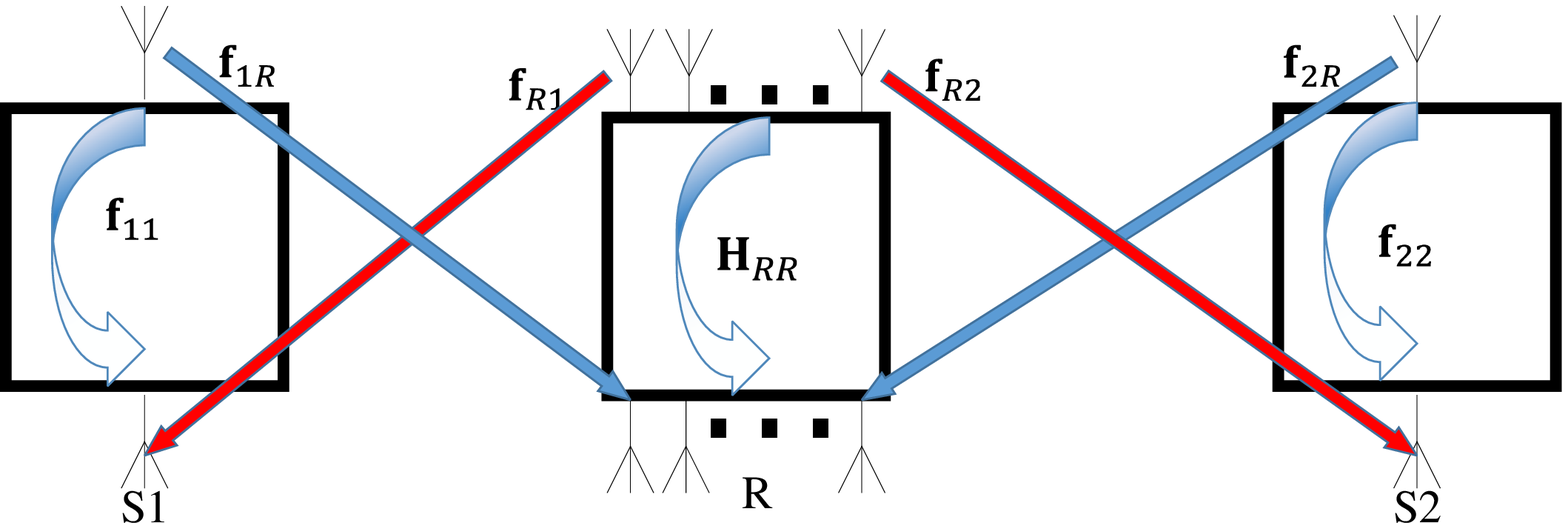}
  \caption{Illustration of the full-duplex MIMO two-way relay network.}
\end{figure}
\section{Problem Formulation}\label{problem}
In this section, the goal is to find the best $\mathbf{w}$ such that the minimum rate of $S_i$ is maximized, when the node $R$ is subject to the power constraint ${{P}_{R}}$. The optimization problem can be written as follows
\begin{equation}\label{opt_prob}
\begin{aligned}
\underset{\mathbf{w}}{\mathop{Max}}\,\,\,&Min\,\left\{ \frac{{{\mathbf{w}}^{H}}{{\mathbf{H}}_{1}}\mathbf{w}}{{{\mathbf{w}}^{H}}{{\mathbf{F}}_{1}}\mathbf{w}+{{\phi }_{1}}},\frac{{{\mathbf{w}}^{H}}{{\mathbf{H}}_{2}}\mathbf{w}}{{{\mathbf{w}}^{H}}{{\mathbf{F}}_{2}}\mathbf{w}+{{\phi }_{2}}} \right\}\\
&s.t.\,\,\,\,\,\,{{\mathbf{w}}^{H}}{{\mathbf{C}}_{1}}\mathbf{w}\le {{P}_{R}}\\
&~~~~~~\,{{\mathbf{w}}^{H}}{{\mathbf{C}}_{2}}\mathbf{w}=0.
\end{aligned}
\end{equation}
Note that in (\ref{opt_prob}) the logarithm function due to the monotonicity is omitted. Also, the second constraint is the ZFC.
Employing the SDR technique, the equality $\mathbf{G}=\mathbf{w}{{\mathbf{w}}^{H}}$ is defined. Also, defining
\begin{equation}\label{define_j}
j=Min\,\left\{ \frac{{{\mathbf{w}}^{H}}{{\mathbf{H}}_{1}}\mathbf{w}}{{{\mathbf{w}}^{H}}{{\mathbf{F}}_{1}}\mathbf{w}+{{\phi }_{1}}},\frac{{{\mathbf{w}}^{H}}{{\mathbf{H}}_{2}}\mathbf{w}}{{{\mathbf{w}}^{H}}{{\mathbf{F}}_{2}}\mathbf{w}+{{\phi }_{2}}} \right\},
\end{equation}
and after some mathematics, (\ref{opt_prob}) can be converted to
\begin{equation}\label{opt_problem_convert}
\begin{aligned}
&\underset{\mathbf{G},\,\,j}{\mathop{Max}}\,\,\,j\\
&s.t.~~Tr\left( \left( {{\mathbf{H}}_{i}}-j{{\mathbf{F}}_{i}} \right)\mathbf{G} \right)\ge j{{\phi }_{i}};\,\,i\in 1,2\\
&~~~~~\,\,Tr\left( {{\mathbf{C}}_{1}}\mathbf{G} \right)\le {{P}_{R}}\\
&~~~~~\,\,Tr\left( {{\mathbf{C}}_{2}}\mathbf{G} \right)=0\\
&~~~~~\,\,\mathbf{G}\underline{\succ }0,\,\,Rank\left( \mathbf{G} \right)=1,
\end{aligned}
\end{equation}
where the rank one constraint is non-convex and, according to the SDR technique, it should be dropped. Moreover, to tackle the problem, for any known value of $j$, the following feasibility problem should be solved.
\begin{equation}\label{opt_feasibility_problem}
\begin{aligned}
&Find~\mathbf{G}\\
&s.t.~~Tr\left( \left( {{\mathbf{H}}_{i}}-j{{\mathbf{F}}_{i}} \right)\mathbf{G} \right)\ge j{{\phi }_{i}};\,\,i\in 1,2\\
&~~~~~\,\,Tr\left( {{\mathbf{C}}_{1}}\mathbf{G} \right)\le {{P}_{R}}\\
&~~~~~\,\,Tr\left( {{\mathbf{C}}_{2}}\mathbf{G} \right)=0\\
&~~~~~\,\,\mathbf{G}\underline{\succ }0.
\end{aligned}
\end{equation}
The problem (\ref{opt_feasibility_problem}) is in the form of a SDP problem which can be solved using optimization packages such as CVX which is employed in the current study~\cite{cvxx}. In fact, knowing $j$, the problem in (\ref{opt_feasibility_problem}) is abstracted to finding a feasible matrix $\mathbf{G}$ which meets all the constraints. Finally, the maximum value of $j$, i.e., $j_{Max}$, is sought in the interval $[0,j_{up}]$ where the value of $j_{up}$ is addressed in the Appendix.

Accordingly, a one-dimensional search is carried out to find the maximum value of $j$ using the so-called bisection method as is presented in Algorithm 1.
\begin{algorithm}[t]
 \caption{Bisection search algorithm to find $j_{Max}$}
 \begin{algorithmic}[1]
  \STATE Consider the interval $[0,j_{up}]$.
  \STATE Define $j_1=0$ and $j_2=j_{up}$.
  \STATE Set $j=\frac{j_1+j_2}{2}$.
  \STATE Solve the problem (\ref{opt_feasibility_problem}).
  \STATE Problem (\ref{opt_feasibility_problem}) is feasible, set $j_1=j$; otherwise,   set $j_2=j$.
 \STATE If $j_2-j_1<10^{-4}$, $j_{Max}=j_1$; otherwise, go to step 2
 \end{algorithmic}
 \end{algorithm}
Knowing $j_{Max}$, the corresponding matrix $\mathbf{G}$, i.e. $\mathbf{G}^*$, can be identified. If $\mathbf{G}^*$ is of rank one, the optimal beamforming vector $\bold{w}^*$ can be readily computed from the principal eigenvector of $\mathbf{G}^*$. However, there is no guarantee that $\mathbf{G}^*$ is of rank one; therefore, $j_{Max}$ is considered as the upper bound solution of (\ref{opt_prob}) and the principal eigenvector of $\mathbf{G}^*$ gives the lower bound solution of (\ref{opt_prob}) if it satisfies the constraints as well; otherwise, a combination of eigenvectors may give a proper lower bound solution.

\section{Numerical Results}\label{simulation}
This section aims to represent the numerical results of the proposed approach and compare the results to that of proposed in~\cite{maxminSINRref} which is being served as the benchmark in the current study. It should be noted that in~\cite{maxminSINRref} the relay operates in HD mode, while the proposed work deals with a FD TW relay and that both works assume multi-antenna nodes.

Throughout the simulations, all channel coefficients are assumed i.i.d. complex zero-mean Gaussian variables of unit power. In addition, in the proposed model, the RSI power is set to -10dB and -40dB to show the impact of RSI on the overall performance of the network, in terms of the minimum rate of the exchange information. Moreover, the noise power of all receiving nodes are set to $0$dBW, i.e. $\sigma _{R}^{2}=\sigma _{i}^{2}=\text{0dBW}$.

To have a fair comparison result, for the current work as well as the work presented in~\cite{maxminSINRref}, a quarter of total power ($P_T$) is assigned to either of transmitting nodes, i.e., $p_i=\frac{P_T}{4}$, and the remaining power is set to the relay, i.e.,~$P_R=\frac{P_T}{2}$. In addition, for each transceiver in~\cite{maxminSINRref}, two antennas are assumed. Also, in all simulations for the current study, half of the relay's antennas are devoted to the reception while the remaining antennas are engaged with transmission. Accordingly, the average results for 1000 channel realizations are being reported in the numerical results.
\begin{figure}[t]
\centering
 \includegraphics[scale=.5]{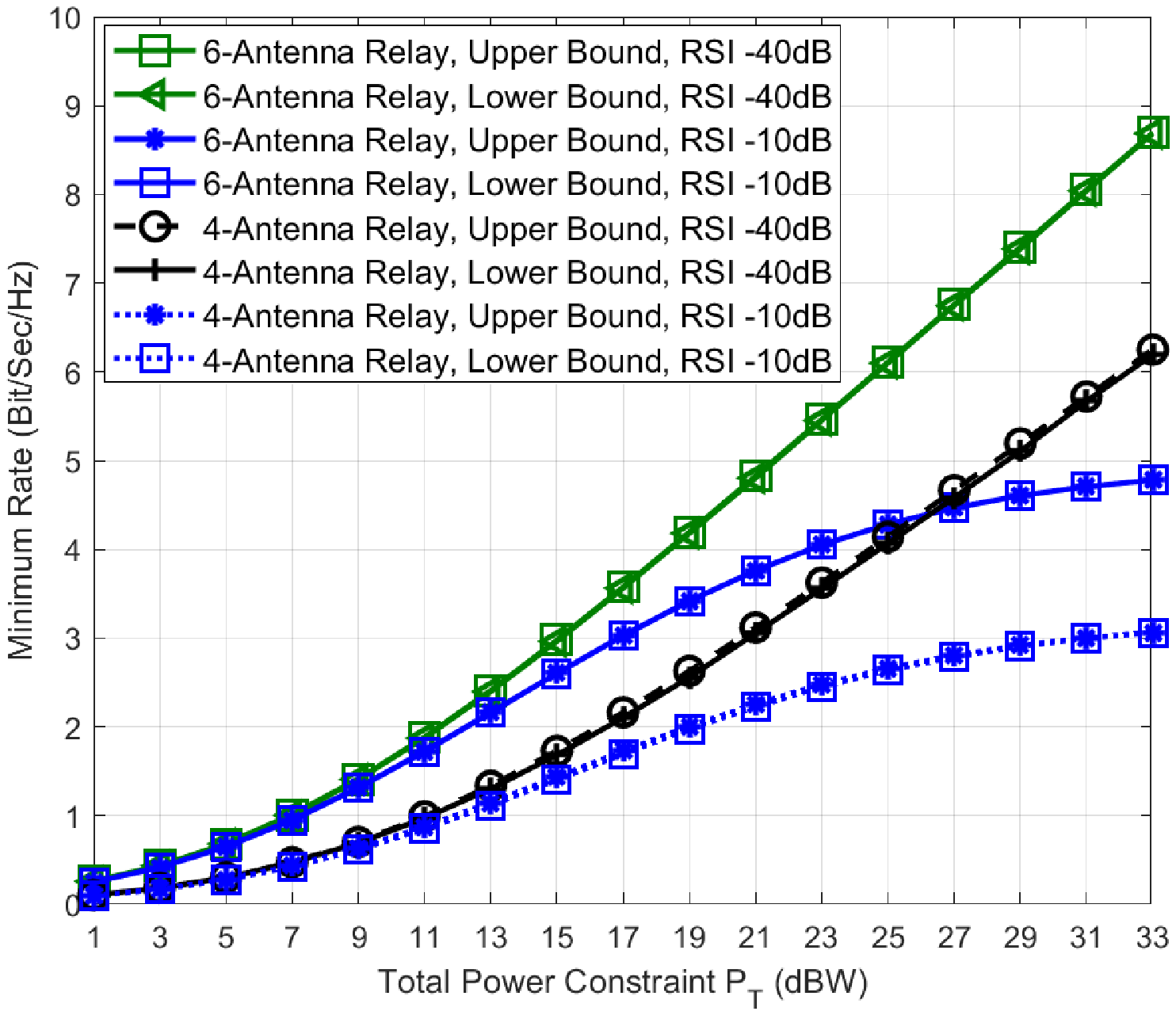}
  \caption{The average minimum rate vs. the total power constraint in a network with a 4- and 6-antenna relay.}\label{Fig1}
\end{figure}
\begin{figure}[t]
\centering
 \includegraphics[scale=.5]{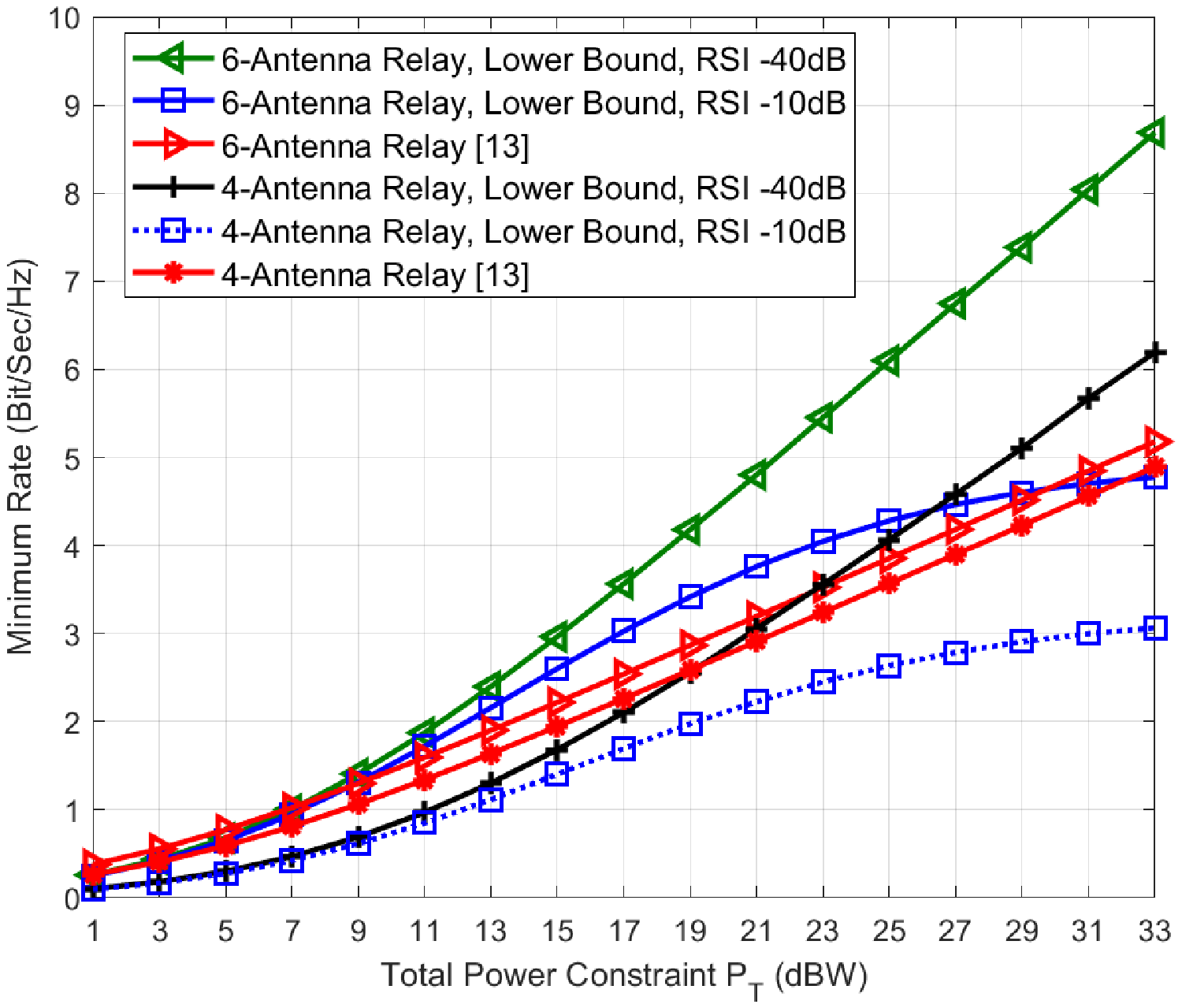}
  \caption{The average minimum rate vs. the total power constraint in a network with a 4- and 6-antenna relay.}\label{Fig2}
\end{figure}

Fig. \ref{Fig1} is provided to show the proposed upper and lower bound solutions of the underlying problem for cases that the network uses a relay equipped with 4 or 6 antennas. As illustrated, upper and lower bound curves for each case closely follow each other, showing that the proposed approach can be regarded as a close-to-optimal solution.

Fig. \ref{Fig2} shows the proposed lower bound~(achievable) solution and the average minimum rate obtained by the presented approach in~\cite{maxminSINRref} for a network using a relay equipped with 4 or 6 antennas. In fact, Fig. \ref{Fig2} compares the minimum achievable rate of FD and HD networks. As seen, the performance of the proposed approach is degraded when RSI takes the value -10dB. However, for the case that RSI is sufficiently low, i.e., when RSI is -40dB, by increasing the number of antennas at the relay, the superior performance of the proposed approach compared to the HD scenario becomes more conspicuous.

\section{Complexity Analysis}\label{complexity}
In this section, we are going to compare the computational complexity of the proposed approach with that of presented in~\cite{maxminSINRref}. Note that for a SDR problem whose unknown matrix is $q\times q$ and the number of linear constraints is $p$, the worst-case complexity is given by $\mathcal{O}\left( \max {{\left\{ p,q \right\}}^{4}}{{q}^{\frac{1}{2}}}\log \left( 1/\varepsilon  \right) \right)$, where $\varepsilon > 0$ is the solution accuracy \cite{sroqop}. Since for the current study as well as the method presented in~\cite{maxminSINRref}, the SDR technique is employed, the complexity can be computed as Table \ref{table:1}. In this table, $l_1$ and $l_2$ indicate the number of searches required for the convergence of the algorithms proposed in the current work and presented in~\cite{maxminSINRref}, respectively. In addition, for the sake of fair comparison, the number of $k$ antennas at the relay is assumed the same for both studies.
Considering large values of $k$ and equal number of search steps for both cases, i.e., $l_1=l_2=l$, the complexity of the proposed approach and that of given in~\cite{maxminSINRref} become $\mathcal{O}(l{k^3})$.
\begin{table}
\centering
\caption{The complexity analysis}\label{table:1}
\begin{tabular}{ |c|c| }
\hline
 Problem & Complexity  \\
\hline
 Proposed Approach & $\mathcal{O}\left(l_1 Max \left\{ 4,\frac{k^2}{4} \right\}\frac{k}{2}\log \left( 1/\varepsilon  \right) \right)$  \\
   \hline
 The problem in~\cite{maxminSINRref}  & $\mathcal{O}\left(l_2 Max \left\{ 3,k^2+1 \right\}(k^2+1)^{\frac{1}{2}}\log \left( 1/\varepsilon  \right) \right)$  \\
 \hline
\end{tabular}

\end{table}

\section{Conclusion}\label{conclusion}
In this paper, the problem of maximizing the minimum rate of a relay-assisted multi-antenna FD TW network has been investigated. Applying a ZFC to the beamforming matrix of the relay to suppress RSI, upper and lower bound solutions of the problem have been proposed. Numerical results demonstrate that the proposed upper and lower bound solutions closely follow each other; therefore, the proposed approach yields a close-to-optimal solution. Also, it is shown that the proposed method for small values of RSI power, exhibits better performance as compared with the method proposed in~\cite{maxminSINRref} for HD networks which to the best of authors' knowledge is the best-known method addressed in the literature.

\appendix[Upper and Lower bounds for $j$]
In this appendix, details of finding the upper and lower bounds of $j$ is discussed. As mentioned earlier, $j$ is defined by (\ref{define_j}). The numerators of the fractional terms of (\ref{define_j}) are power values; thus, they are non-negative values. Mathematically speaking, the lowest value of both terms ${{{\mathbf{w}}^{H}}{{\mathbf{H}}_{1}}\mathbf{w}}$ and ${{{\mathbf{w}}^{H}}{{\mathbf{H}}_{2}}\mathbf{w}}$ is zero; therefore, the lower bound for $j$ is zero.

To find the upper bound of $j$, the two fractional terms inside of the minimum function in (\ref{define_j}) should be maximized and the greater value between the two maximized values can be selected as the upper bound of $j$, i.e. $j_{up}$ . Hence, the following optimization problem must be solved for $i\in \{1,\,2\}$, and the cost function which has the greater value is chosen as $j_{up}$.
\begin{equation}\label{upperbound_j_opt}
\begin{aligned}
&\underset{\mathbf{w}}{\mathop{Max}}\,\,\,\frac{{{\mathbf{w}}^{H}}{{\mathbf{H}}_{i}}\mathbf{w}}{{{\mathbf{w}}^{H}}{{\mathbf{F}}_{i}}\mathbf{w}+{{\phi }_{i}}}\\
&s.t.~~\,\,\,\,{{\mathbf{w}}^{H}}{{\mathbf{C}}_{1}}\mathbf{w}\le {{P}_{R}}\\
&~~~~~~~\,{{\mathbf{w}}^{H}}{{\mathbf{C}}_{2}}\mathbf{w}=0.
\end{aligned}
\end{equation}
After applying the SDR technique and the definition ${{\mathbf{V}}_{i}}=\mathbf{w}{{\mathbf{w}}^{H}}$, (\ref{upperbound_j_opt}) is transformed into
\begin{equation}\label{upperbound_j_optSDR}
\begin{aligned}
&\underset{{{\mathbf{V}}_{i}}}{\mathop{Max}}\,\,\,\frac{Tr\left( {{\mathbf{H}}_{i}}{{\mathbf{V}}_{i}} \right)}{Tr\left( {{\mathbf{F}}_{i}}\mathbf{V} \right)+{{\phi }_{i}}}\\
&s.t.~~\,\,\,\,Tr\left( {{\mathbf{C}}_{1}}{{\mathbf{V}}_{i}} \right)\le {{P}_{R}}\\
&~~~~~~~\,Tr\left( {{\mathbf{C}}_{2}}{{\mathbf{V}}_{i}} \right)=0\\
&~~~~~~~\,{{\mathbf{V}}_{i}}\underline{\succ }0,\,\,Rank\left( {{\mathbf{V}}_{i}} \right)=1,
\end{aligned}
\end{equation}
where according to the SDR technique, the non-convex rank-one constraint should be omitted.
Also, (\ref{upperbound_j_optSDR}) can be simplified using the Charnes-Cooper transformation~\cite{charnescooper}. According to this transformation, we define ${{\mathbf{V}}_{i}}=\frac{{{\mathbf{Y}}_{i}}}{\omega_i }$, where the scalar $\omega_i>0$ and the matrix $\mathbf{Y}_i$ should be found such that the denominator of the objective function of (\ref{upperbound_j_optSDR}) times $\omega_i$ equals the unit value. Therefore, (\ref{upperbound_j_optSDR}) can be rewritten as follows
\begin{equation}\label{upperbound_j_optCharness}
\begin{aligned}
&\underset{\mathbf{Y}_i,\,\,\omega_i}{\mathop{Max}}\,\,\,Tr\left( {{\mathbf{H}}_{i}}{{\mathbf{Y}}_{i}} \right)\\
&s.t.~~\,\,\,\,Tr\left( {{\mathbf{C}}_{1}}{{\mathbf{Y}}_{i}} \right)\le \omega_i {{P}_{R}}\\
&~~~~~~~\,Tr\left( {{\mathbf{C}}_{2\,}}{{\mathbf{Y}}_{i}} \right)=0\\
&~~~~~~~\,Tr\left( {{\mathbf{F}}_{i}}{{\mathbf{Y}}_{i}} \right)+\omega_i {{\phi }_{i}}=1\\
&~~~~~~~\,\omega_i >0,\,\,{{\mathbf{Y}}_{i}}\underline{\succ }0.
\end{aligned}
\end{equation}
As seen, (\ref{upperbound_j_optCharness}) is a SDP problem which can be solved by CVX package.

It is also worth noting that since finding the upper bound is of interest here, if the obtained matrix ${{\mathbf{Y}}_{i}}$, and consequently ${{\mathbf{V}}_{i}}$, by solving (\ref{upperbound_j_optCharness}) is not of rank one, the obtained value of the cost function of (\ref{upperbound_j_optCharness}) becomes greater than the case where ${{\mathbf{Y}}_{i}}$ is of rank one. This is due to the fact that when the obtained matrix is not of rank one the cost function of (\ref{upperbound_j_optCharness}) is the upper bound solution, and this has no effect on the result of the Algorithm 1 as it merely gives a greater upper bound of $j$.

\end{document}